\documentstyle[prb,twocolumn,aps,epsf]{revtex}
\begin{document}
\draft
\wideabs{
\title{Diffusion and Transport Coefficients in Synthetic Opals}
\author{J. O. Sofo}
\address{
Centro At\'{o}mico Bariloche and Instituto Balseiro, \\
Comisi\'{o}n Nacional de Energ\'{\i}a At\'{o}mica, (8400) Bariloche RN, Argentina}
\author{G. D. Mahan}
\address{
Solid State Division, Oak Ridge National Laboratory,
P.O.Box 2008, Oak Ridge, TN 37831-6030,
and\\
Department of Physics and Astronomy, The University of Tennessee,
Knoxville, TN 37996-1200}
\date{\today}
\maketitle 
\begin{abstract}
Opals are structures composed of the closed packing of spheres in the
size range of nano-to-micro meter. They are sintered to create small
necks at the points of contact. We have solved the diffusion problem
in such structures. The relation between the diffusion coefficient and
the termal and electrical conductivity makes possible to estimate the
transport coefficients of opal structures. We estimate this changes as
function of the neck size and the mean-free path of the carriers.
The theory presented is also applicable to the diffusion problem in
other periodic structures.
\end{abstract}
\pacs{72.20.-i, 66.10.Cb, 72.15.Jf}
}
\narrowtext

\section{Nanostructured thermoelectrics}

A good thermoelectric material has low thermal conductivity
$\kappa$, high electrical conductivity $\sigma$, and high Seebeck
coefficient, in order to maximize the thermoelectric figure of merit
\begin{equation}
Z=\frac{\sigma S^2}{\kappa}\;.
\end{equation}
$Z$ has units of inverse temperature and is generally quoted as $ZT$,
with $T$ the absolute temperature. For more than 40 years, the search
for better thermoelectrics has not provided a material with $ZT$
significantly larger than one. $ZT$ of about 4 would make
thermoelectric coolers capable to compete against the currently
established devices based on gas-compression technology.

A large effort has been done so far in order to improve the efficiency
of thermoelectric materials. \cite{tritt,disalvo} The search for
better thermoelectrics can be classified by its length scale. The
improvement was attempted in the microscopic, mesoscopic and
macroscopic scale. In the microscopic scale, most of the search has
been done by studying variations of the chemical composition of
prospect materials and alloys. On the opposite range, the macroscopic
range involves the design of the thermoelectric devices. So far, most
of the effort has been done in these length scales. Only recently, with
the improvement of different nano-fabrication techniques, the search
for better thermoelectrics has been done in the mesoscopic length
scale. This includes the study of superlattices, quantum dots and
opals. It is interesting to mention an essential difference between
this and the previously mentioned length scales. The theoretical tools
needed to evaluate the improvements in the micro and macro length
scales are fairly well established. Both, the transport properties of
compounds and alloys, and the macroscopic equations for device
modeling are well known. However, this is not the case in the
mesoscopic scale. In many possible structures, we are still working on
the basic theoretical framework to understand and plan the new
designs. A clear example of this is the discussion around the
thermoelectric application of superlattices.\cite{super} The purpose
of the present work is to develope theoretical tools to evaluate
the transport coefficients of one of these mesoscopic structures: the
synthetic opals.

Synthetic opals are nanostructured materials composed of closed-packed
spheres with a uniform radius $a$ which can range from nanometers to
microns. They are often made with glass spheres, but various
techniques can be used to replace the glass spheres with almost any
solid state material. The opals are usually sintered so that the
points of contact between neighboring spheres become small
necks. Currents of electricity or heat can go from sphere to sphere
through these necks. Here we calculate the electrical and thermal
conductivity of such opal structures. We ask the question: how does
the opal structure reduce the electrical and thermal conductivity?
Assuming that the Seebeck coefficient will not be affected by the opal
structure, if the thermal conductivity is reduced much more than the
electrical conductivity, the opals could be useful thermoelectric
materials.

Currently, there is an experimental group at AlliedSignal Corporation
preparing and developing these structures to evaluate their potential
as thermoelectrics. Our theory provides a theoretical tool to guide
this evaluation and design.\cite{Anvar1,Anvar2} The thermal
conductivity of nanocomposites with regular structure have been
studied by B.\ N.\ Bogomolov and collaborators and is reported is a
series of publication from 1995 to date.
\cite{bogo1,bogo2,bogo3,bogo4,bogo5,bogo6} The object of these studies
has been a SiO$_2$ nanocomposite opal with first order voids
completely filled with NaCl. Their results are analyzed with a theory
for composite materials developed by R. E. Meredith and C. W. Tobias
\cite{meredith60} on the basis of the original work of Lord Rayleigh
\cite{rayleigh1892}. The theory is based in the solution of the Laplace
equation for a regular array of spheres embedded in a different
medium. This theory can not be applied to the case for which the spheres
are touching each other, which is the focus of our present interest. 

In Sec.~\ref{tcad} we relate the diffusion coefficient in the opal with
the transport coefficients that are relevant for a thermoelectric
analysis. The diffusion coefficient is evaluated with a random-walk
model as explained in Sec.~\ref{dcarw} and this model is
analyzed and solved in Secs.~\ref{dco} and \ref{prop}. The results of
applying this model to an fcc opal structure are discussed in
Sec.~\ref{fcc}.

\section{Transport coefficients and diffusion}
\label{tcad}

Given a bulk material with thermal conductivity $\kappa$ and
electrical conductivity $\sigma$, we want to estimate the thermal and
electrical conductivities, $\kappa^{(op)}$ and $\sigma^{(op)}$, of a
synthetic opal made of spheres of this material. The origin and
magnitude of the reduction in the transport coefficients depends on
the mean free path of the carriers $\ell$ compared with the diameter
of the spheres $d$. If $\ell<<d$ the effect is due to a change in the
boudary conditions, i.e. the thermal conductivity of the overall
structure can be obtained by solving Laplace equation in the opal
geometry. If $\ell\approx d$ the boundary scattering produced by the
sphere surface and the necks becomes important. In addition to the
geometrical effect, which is also present in this case, there is a
reduction of the intrinsic conductivities due to the presence of the
boundaries.  In order to cover both situations with our model
calculation, we will relate the transport coefficients to the diffusion
coefficient of the opal and calculate this diffusion coefficient
as a function of the mean free path and the opal geometric parameters,
which are the diameter of the spheres $d$ and the distance between its
centers $a$.

The lattice thermal conductivity is proportional to the diffusion
coefficient for phonons
\begin{equation}
\kappa=CD_{ph}\;,
\label{kappa}
\end{equation}
where $C$ is the specific heat of the material.  Since we are mostly
interested in semiconductors at room temperature, it is accurate to
use the Einstein's relation \cite{landau} to relate the mobility
to the diffusion coefficient for electrical carriers
\begin{equation}
\mu=\frac{e D_{el}}{k_BT}\;,
\label{mu}
\end{equation}
where $e$ is the electron charge and $k_B$ is the Boltzmann constant. 

With Eqs.~(\ref{kappa}) and (\ref{mu}) in mind, to estimate the
themoelectric figure of merit of opals we have to solve the diffusion
problem in the opal structure. In this way we will obtain a relation
between the diffusion constant $D$ of the bulk material and the
overall diffusion constant of the opal $D^{(op)}$. The
thermoelectric figure of merit $Z$ is proportional to $D_{el}/D_{ph}$.
To evaluate the improvement produced by the opal we have to compare
this ratio in bulk and in the opal.

\section{Diffusion Coefficient and Random Walks}
\label{dcarw}

Consider a particle moving with velocity $v$ between collisions, each
collision randomly changing the direction of the movement. In this situation
the diffusion coefficient is given by
\begin{equation}
D=\lim_{t\rightarrow\infty}\frac{\langle r(t)^2 \rangle }{6t}\;,
\label{difu}
\end{equation}
where $\langle r(t)^2 \rangle$ is the mean-square displacement of the
particle and $t$ the total time of the random walk.
This can be calculated by running a random walker that performs a step
of length $\ell$ at every time step of length $\tau$.
S. Chandrasekar \cite{Chandra} solved the diffusion problem of a random
walker in an homogeneous media and obtained 
\begin{equation}
\langle r^2 \rangle = \ell^2 N\;,
\label{r2free}
\end{equation}
where N is the number of steps. By replacing this expression for
$\langle r^2\rangle$ in Eq.~(\ref{difu}) for the diffusion
coefficient, we obtain 
\begin{equation}
D=\frac{1}{6}v\ell\;,
\label{difuvl}
\end{equation}
where $v$ is the velocity of the random walker, i.e. the ratio
$\ell/\tau$. 

The expression given in Eq.~(\ref{r2free}) for the mean-square
displacement is only valid in bulk.  If the carrier moves inside the
spheres of the opal structure we have to calculate $\langle
r^2\rangle$ in a different way in order to use Eq.~(\ref{difu}) and
obtain the diffusion coefficient of the opal $D^{(op)}$.

\section{Diffusion Coefficient in Opals}
\label{dco}

We consider the carriers as classical particles performing a diffusive
motion inside the material that forms the opal. The particle will move
inside a sphere crossing from time to time the neck to a neighboring
sphere. If the length of the jumps is smaller than the diameter of the
balls, the diffusion coefficient inside a ball will be that of the
bulk material with a microstructure corresponding to grains of the
size of the spheres.  The effective diffusion coefficient of the opal
structure will be determined by the diffusion from sphere to sphere.

One method to evaluate this diffusion coefficient is the numerical
simulation of the walker in the opal structure. However, if the
diameter of the necks is small, as it is the case for the experimental
situation, the walker will spend most of the time wandering inside a
sphere and only sporadically crossing to the next sphere. As a
consequence, most of the computer time will be lost in a diffusive
movement inside the spheres and the diffusion coefficient calculation,
which is essentially related with this motion from sphere to sphere,
will become extremely expensive in computer time. A different method
is needed.

Here we present such a method. The central idea is to integrate out
the motion of the carrier inside the sphere while keeping attention on
the motion from sphere to sphere.  The diffusion in opals, with this
picture in mind, is similar to the diffusion of a particle in a
lattice, the lattice made up by the spheres.  The standard theory of
diffusion in a lattice is due to Chudley and Elliot.\cite{ce} However,
their theory was for a particle diffusing from site to site whithout
spending any time in the site. The diffusion in an opal is different
and the theory must include the feature that it takes time to diffuse
within the spheres, from neck-to-neck. The opals are on a lattice, but
the theory must include the diffusion within the spheres.

Define $\gamma_{ij}(t)$ as the probability per unit time that a
particle departs through the neck $i$ at time $t$ if it entered the
sphere through neck $j$ at $t=0$.  Define $f_i({\bf R}_n,t)$ as the
flux of particles departing a sphere centered at ${\bf R}_n$ through
the neck labeled $i$ at time $t$. Note that this quantity is {\it not}
the net flux, which is the number of particles leaving minus the
number entering. It is just the number leaving.  The number entering
the sphere through that neck is included as the flux leaving the
neighboring sphere at ${\bf R}_n+\mbox{\boldmath$\delta$}_i$ through
neck $\bar{i}$. Here we define the conjugate to neck $i$ as
$\bar{i}$. It is the label of the corresponding neck on the
neighboring sphere. A neck $i$ on one sphere is connected to $\bar{i}$
on the neighbor.

The rate equations for particle motion are
\begin{eqnarray}
f_i({\bf R}_n,t)=&\sum_j& \int_0^t dt'\gamma_{ij}(t-t')
f_{\bar{j}}({\bf R}_n+\mbox{\boldmath$\delta$}_j,t')\nonumber\\
&+& \delta_{i,1}\delta_n\delta(t)
\label{equrt}
\end{eqnarray}
The first term on the right is the outgoing flux due to all of the
incoming particles at an earlier time. The last term on the right is
the source term which starts the particle diffusion. The time integral
can be eliminated by a Laplace transform 
\begin{eqnarray}
F_i({\bf R}_n,p)&=&
\int_0^{\infty}dt\: e^{-pt}f_i({\bf R}_n,t)\;,\nonumber\\ 
\Gamma_{ij}(p)&=&
\int_0^{\infty}dt\: e^{-pt}\:\gamma_{ij}(t)\;.
\label{laplace}
\end{eqnarray}
After transforming Eq.~(\ref{equrt}) that describes the particle
motion, it becomes
\begin{eqnarray}
F_i({\bf R}_n,p)&=& \sum_j
\Gamma_{ij}(p)F_{\bar{j}}({\bf R}_n+\mbox{\boldmath$\delta$}_j,p)+\delta_n 
\end{eqnarray}
The lattice properties are taken into account by taking a Fourier transform 
\begin{equation}
G_i({\bf k},p)=
\sum_{{\bf R}_n}e^{i{\bf k}\cdot{\bf R}_n}F_i({\bf R}_n,p)\;,
\label{fourier}
\end{equation}
which leads to the final form of the equation of motion

\begin{equation}
G_i({\bf k},p)=\sum_j \Gamma_{i\bar{j}}G_j({\bf k},p) e^{-i{\bf
k}\cdot\mbox{\boldmath$\delta$}_j}+1
\label{equkp} 
\end{equation}
where we have used
$\mbox{\boldmath$\delta$}_j=-\mbox{\boldmath$\delta$}_{\bar{j}}$.

Eq.~(\ref{equkp})  is a matrix equation which is solved for the
functions $G_i({\bf k},p)$. However, the diffusion properties are
determined by the properties of the matrix 
\begin{equation}
M_{ij} = \delta_{ij}-
\Gamma_{i\bar{j}}\exp({-i{\bf k}\cdot\mbox{\boldmath$\delta$}_j})\;.
\end{equation}
The poles of $G_i$ are determined by the zeros of the determinant of
this matrix. In taking the inverse Laplace transform, the time
dependence is given by the poles of $G_i$, which again are the zeros
of $\det|M_{ij}|$. The smallest pole is defined as that with the
smallest value of $p$. The long time behavior (diffusion) is
determined by the smallest pole.  We now discuss some properties of
the matrix $M$, and the propagators $\gamma$ that will be helpful in
obtaining a form for the diffusion coefficient in opals $D^{(op)}$:

\noindent {\it (i) If there is no absorption, then all of the particles which
enter the sphere eventually must leave it sometime}.
\begin{equation}
1 = \sum_j \int_0^{\infty}\gamma_{ij}(t)= \sum_j \Gamma_{ij}(0)
\label{identity}
\end{equation}
Symmetry shows that this result is independent of $i$. If we define 
\begin{equation}
\Gamma_T(p)=\sum_j \Gamma_{ij}(p)\;,
\end{equation}
the identity in Eq.~(\ref{identity}), takes the form
\begin{equation}
\Gamma_T(0)=1\;.
\label{equal1}
\end{equation}

\noindent {\it (ii) If ${\bf k}=0$ and $p=0$, then $\det|M|=0$.}  If ${\bf
k}=0$ each row of the matrix has the same elements, but in different
order. As a consequence, the eigenvector $\psi_0$ which has all
elements equal to one has the feature that
\begin{equation}
M\psi_0 = (1-\Gamma_T(p))\psi_0\;.
\end{equation}
Therefore at $p=0$ then $M\psi_0=0$. The vector $\psi_0$ has an
eigenvalue of zero at $p=0$. Since the determinant is the product of
the eigenvalues, this also means that the determinant of $M_{ij}$ is
zero in this case.  

\noindent{\it (iii) At ${\bf k}=0$  $\det|M|$ has a factor $1-\Gamma_T(p)$}.
This is a direct consequence of the previous property.

All of the above results are at ${\bf k}=0$. For diffusion one wants
large values of ${\bf R}_n$ which means small values of ${\bf k}$. The
expansion of the determinant at small $p$ and small $k=|{\bf k}|$ has
the general form
\begin{equation}
\det|M|= pu(p)U(p)+ (ka)^2 V(p)\;,
\end{equation}
where $U(p)$ and $V(p)$ are polynomials in the propagators
$\Gamma_{i,j}(p)$, $a$ is the nearest-neighbor distance, and
$u(p)=(1-\Gamma_T(p))/p$.  As a result, the diffusion coefficient of
the opal is given by
\begin{equation}
D^{(op)}= \frac{a^2}{u(0)} \frac{V(0)}{U(0)}\;,
\end{equation}
where $u(0)$ is the limit of $u(p)$ for $p\rightarrow 0$.

These general results are illustrated for some relevant common lattices. 
For each particular lattice,
the propagator $\Gamma_{i,j}$ depends only on the angle between necks
$i$ and $j$ and we will use the notation $\Gamma_\theta$, where
$\theta$ is the angle between the normals to the neck surfaces. 

\subsubsection*{Linear Chain}

We first present the simple case of the linear chain in order to
illustrate the method. In this case, each ball has two necks. We will
define $\Gamma^{(lc)}_0(p)=\Gamma_{11}(p)=\Gamma_{22}(p)$ as the
Laplace transform of the probability that the walker enters and leaves
through the same neck, and $\Gamma^{(lc)}_{180}(p) = \Gamma_{12}(p) =
\Gamma_{21}(p)$ corresponding to the probability of the walkers
leaving the sphere through the opposite side.
In this case the matrix $M$ has the form
\begin{equation}
M^{(lc)}= 
\left[
\begin{array}{cc} 
1-\Gamma_{180}^{(lc)}e^{ika} &
-\Gamma_{0}^{(lc)} e^{-ika}\\
-\Gamma_{0}^{(lc)} e^{ika} &
1-\Gamma_{180}^{(lc)} e^{-ika} 
\end{array}
\right]\;.
\end{equation}
The diffusion coefficient for this one dimensional opal will be 
\begin{equation}
D^{(op)} = \frac{a^2}{2u(0)}\:
\frac{\Gamma_{180}^{(lc)}}{\Gamma_{0}^{(lc)}}\;.
\label{doplc}
\end{equation}

\subsubsection*{Square Lattice}

In the case of the two dimensional square lattice we have three
different possibilities, the walker can go back through the same entry
neck, can go out through one of the two necks that form an angle of 90
degrees with the incoming trajectory, or can leave the sphere through
the opposite neck. The first possibility is represented by
$\Gamma_{0}^{(sq)}(p)$, the second by $\Gamma_{90}^{(sq)}(p)$ and the
last by $\Gamma_{180}^{(sq)}(p)$. In this case 
\begin{equation}
\Gamma_T^{(sq)}(p)=\Gamma_{0}^{(sq)}(p) + 2\Gamma_{90}^{(sq)}(p) +
\Gamma_{180}^{(sq)}(p)\;,
\end{equation}
and the diffusion coefficient is given by 
\begin{equation}
D^{(op)} = \frac{a^2}{4u(0)}\:
\frac{\Gamma_{180}^{(sq)}+\Gamma_{90}^{(sq)}}
     {\Gamma_{90}^{(sq)}+\Gamma_{0}^{(sq)}}
\;.
\label{dopsq}
\end{equation}

\subsubsection*{Simple Cubic Lattice}

The simple cubic lattice has four necks at an angle of 90 degrees, as
a consequence
\begin{equation}
\Gamma_T^{(sc)}(p)=\Gamma_{0}^{(sc)}(p) + 4\Gamma_{90}^{(sc)}(p) +
\Gamma_{180}^{(sc)}(p)\;,
\end{equation}
and the diffusion coefficient is given by 
\begin{equation}
D^{(op)} = \frac{a^2}{6u(0)}\:
\frac{\Gamma_{180}^{(sq)}+2\Gamma_{90}^{(sq)}}
     {2\Gamma_{90}^{(sq)}+\Gamma_{0}^{(sq)}}
\;.
\label{dopsc}
\end{equation}

\subsubsection*{Face Centered Cubic Lattice}

The fcc lattice corresponds to the close packing of spheres observed
experimentally in opals. In the fcc lattice each sphere is connected
through twelve necks to its nearest neighbors. Assume that we enter
the sphere through neck 1, in the fcc lattice there are four necks
with a normal forming an angle of 60$^\circ$ with the normal of neck
1, two at 90$^\circ$, four at 120$^\circ$, and one neck in the
opposite side of neck 1 at 180$^\circ$.

The result for the fcc lattice is:
\begin{eqnarray}
\Gamma_T^{(fcc)}(p)&=&\Gamma_{0}^{(fcc)}(p) + 4\Gamma_{60}^{(fcc)}(p)
+ 2\Gamma_{90}^{(fcc)}(p) +\nonumber\\
&+& 4\Gamma_{120}^{(fcc)}(p) + \Gamma_{180}^{(fcc)}(p)
\end{eqnarray}
and
\begin{eqnarray}
 D^{(op)}&=&
\frac{\delta^2}{3 u(0)}\label{dopfcc}\\ &\times&
\frac{\Gamma_{60}^{(fcc)}(0)+\Gamma_{90}^{(fcc)}(0) +
3\Gamma_{120}^{(fcc)}(0)+\Gamma_{180}^{(fcc)}(0)} {\Gamma_{0}^{(fcc)}(0)+3\Gamma_{60}^{(fcc)}(0) +
\Gamma_{90}^{(fcc)}(0)+\Gamma_{120}^{(fcc)}(0)} \nonumber
\end{eqnarray}

The final stage of our calculation of the diffusion coefficient in
opals is the determination of the particle propagators inside the
sphere, $\gamma_{ij}(t)$, and this is the focus of the next section.

\section{Propagators inside the spheres}
\label{prop}

The propagator inside the sphere, $\gamma_{ij}(t)$, was defined as the
probability of a particle leaving through neck $j$ at time $t$ if
entered the sphere through neck $i$ at $t=0$. We obtain this
propagators by numerical simulation of the walker. A specific behavior
of the walker when it hits the surface has to be specified to do the
simulation. We found that the final result is not strongly dependent
on the particular scattering at the surface as long as the particles
are not absorbed. All the results shown here have been performed assuming
that the walker starts a new random step after hitting the wall.

\begin{figure}
\epsfverbosetrue
\epsfxsize=8cm
\epsfysize=5.5cm
\epsfbox{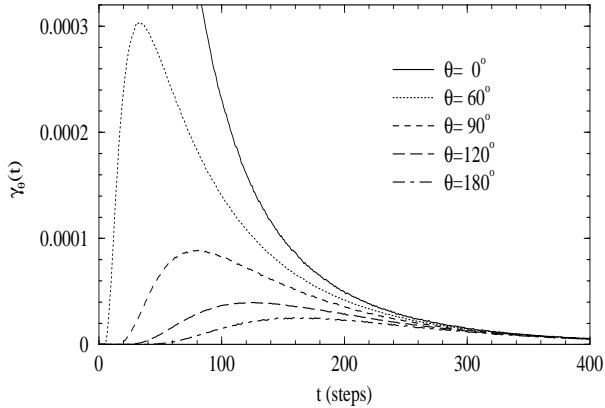}
\caption{Propagators as a function of time of a random walker inside
an opal structure with diameter of the spheres $d$ 5\% larger than the
intercenter spacing $a$. The mean free path $\ell=0.06 a$.}
\label{fig:prop}
\end{figure}

Since it is experimentally observed that the synthetic opal adopts a
closed-packed arrangement of the spheres we will concentrate in the
simulation of an fcc arrangement of spheres.  In Fig.~\ref{fig:prop}
we show the time dependence of the propagators for an opal in which
the diameter of the spheres $d=1.05\:a$ and the step length of the
walker $\ell=0.06\:a$. The propagator for leaving through the same
neck of entry, $\gamma_{0}(t)$, is the only one that does not go to
zero for $t\rightarrow 0$. There are many events in which the walker
spends just a few steps inside the ball and goes back through the same
neck. To leave through other neck the walker needs at least a number
of steps equal to the ratio between the shortest distance between
these necks and the step length. Of course this contribution
corresponds to events with very low probability in which the walker
reaches the other neck in an almost ballistic way. All the propagators
converge to the same exponential decay for long times. It is clear
that after a long time of wandering inside the sphere, the
walker looses all memory of the entry neck.

To calculate the diffusion coefficient we need the Laplace transform
of this functions for $p=0$ that is the integral over all times, and
the function $u$ also evaluated at $p=0$. From its definition we see
that 
\begin{equation}
u(0)=-\left.\frac{d\Gamma_T(p)}{dp}\right)_{p=0}=\int_0^\infty
t\:\gamma_T(t)\:dt\;.
\end{equation}
Both integrals are very easy to obtain from the numerical simulation.

\section{Diffusion Coefficient of a face centered cubic opal}
\label{fcc}

\begin{figure}
\epsfverbosetrue
\epsfxsize=8cm
\epsfysize=5.5cm
\epsfbox{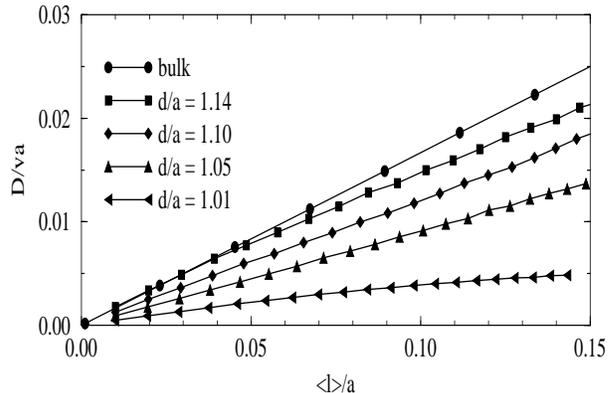}
\caption{ Diffusion coefficient as a function of the mean free path
for different diameters of the spheres. The diffusion coefficient is
expressed in units of the velocity of the carriers times the lattice
constant. The mean free path used for the plot is an average of the
time step. The diffusion coefficient of the bulk material $D=vl/6$ is
show as a reference.}
\label{fig:dop}
\end{figure}

We have used the theory described above to evaluate the diffusion
coefficient of an fcc opal structure.  Figure \ref{fig:dop} shows
the result as a function of the step length, for different sphere
diameters. From the figure is clear the linear behavior of the
diffusion coefficient with the step length, at least when it is
shorter than 15\% the sphere diameter. The diffusion coefficient of
the bulk material is show as a reference.  

The result that the diffusion coefficient of the opal is a lineal
function of the mean free path of the carriers is not promising for
the opals as thermoelectric materials. This linearity implies that the
ratio $D^{(op)}_{el}/D^{(op)}_{ph}$ will be approximately the same as
the ratio $D_{el}/D_{ph}$ and no increase of the thermoelectric figure
of merit is to be expected. 

\begin{figure}
\epsfverbosetrue
\epsfxsize=8cm
\epsfysize=5.5cm
\epsfbox{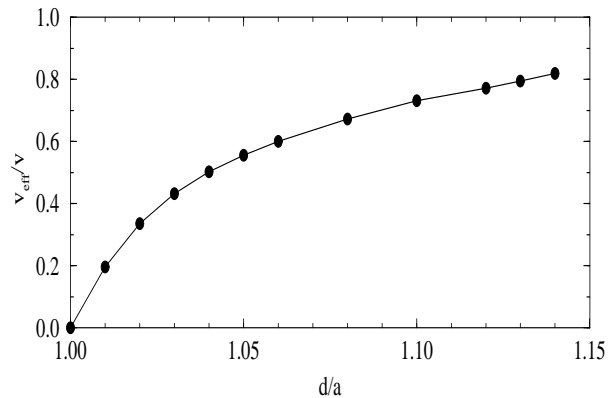}
\caption{Effective velocity as a function of the diameter of the
spheres in the opal.}
\label{fig:slope}
\end{figure}

The slope depends on the diameter of the balls and defines an
effective speed of the carrier.  This dependence is shown in
Fig.~\ref{fig:slope}. This figure shows clearly that the diffusion
coefficient goes to zero if the diameter of the spheres is smaller
than the distance between the centers. This is because we assume a
zero conductivity for the intersphere material. When the diameter is
increased the effective velocity increases approaching that of the bulk
material.

\section{Summary and Conclusions}

We have developed a theory for diffusion of particles in an opal
structure. With this theory, we calculate the diffusion coefficient of
the opal as a function of the diffusion coefficient of the constituent
material, the size of the spheres and the necks between them. We apply
this theory to some of the most common lattices and provide an
explicit expression for the diffusion coefficient in each case.

The experimentally most relevant lattice is the fcc.  The application
of this theory to a synthetic fcc opal structure shows that the
diffusion coefficient is linear with the mean free path of the
carriers. This result holds up to mean free paths of the order of 15\%
the diameter of the spheres. The effect of the opal structure is to
reduce the effective speed of the carriers. We calculate the
dependence of this effective velocity with the size of the spheres.

The linear dependence of the diffusion coefficient with the mean free
path indicates that the overall reduction for electrons and phonons
will be the same. Assuming that the Seebeck coefficient is not
affected by the opal structure, we do not expect any increase in the
thermoelectric figure of merit of opals compared with the bulk
material. A mayor effect of the opal structure is expected in the case
of a mean free path for phonons larger or of the order of the sphere
diameter. However, this will correspond to a material which is a good
thermal conductor in bulk and the reduction should overcome this
unfavorable starting point.

As a final comment, there is an interesting property of all the
expressions obtained for the diffusion coefficient in the different
lattices, Eqs. (\ref{doplc}), (\ref{dopsq}), (\ref{dopsc}), and
(\ref{dopfcc}). In all these cases, adding the numerator and
denominator $\Gamma_T(0)$ is obtained, i.e. the sum is equal to
1. Even though this relation is very suggestive we could not find a
good explanation for it.

The theory developed in this work is applicable, not only to synthetic
opal structures but, can be used to solve the diffusion problem in any
periodic structure in which the walker spends some time inside the
unit cell.

\acknowledgments

We acknowledge research support from the D.O.D. Advanced Research
Projects Agency under contract  DAAB07-97-J036 with the Allied
Signal Corporation. J.O.S. is supported by CONICET, Argentina.
We want to thank Dr. M. Bartkowiak for his comments on our
manuscript.

\end{document}